\documentclass[12pt]{iopart}
\pdfoutput=1
\usepackage{iopams}
\expandafter\let\csname equation*\endcsname\relax
\expandafter\let\csname endequation*\endcsname\relax
\usepackage{amsmath,mathtools}
\usepackage{graphicx}
\usepackage[usenames,dvipsnames]{color}
\usepackage{url}
\usepackage{epstopdf}

\usepackage{cite}

\definecolor{linkcolor}{rgb}{0,0,0.6} 
\usepackage[colorlinks=true, pdfstartview=FitV, linkcolor= linkcolor, citecolor= linkcolor, urlcolor= linkcolor, hyperindex=true,hyperfigures=true]{hyperref}

\renewcommand{\O}{{X}}
\newcommand{\KL}{{\mathbb S}}
\newcommand{\caK}{{\mathcal K}}
\newcommand{\kappab}{{\overline\kappa}}
\newcommand{\sigmab}{{\overline\sigma}}
\newcommand{\Mean}[1]{{\left< {#1} \right>}}
\newcommand{\mean}[1]{{\langle {#1} \rangle}}

\newcommand{\refFT}{eva93,gal95,kur98,jar97,cro99,mae99,leb99,fal15c}
\newcommand{\refFDT}{cug94,lip05,bai09,sei10,ver11,bas15b,yol17}
\newcommand{\refNEI}{bar15,pie16,pie16a,pol16,gin16,gar17,nar17,mae17,pie17,gin17,hor17,pro17,chi18,dec18,DS18,D18}

\begin{document}

\title[Kinetic uncertainty relation]{Kinetic uncertainty relation}

\author{Ivan Di Terlizzi$^{*,\dag}$}
\author{Marco Baiesi$^{*,\dag}$}

\address{$*$ Dipartimento di Fisica e Astronomia ``Galileo Galilei'',
  Universit\`a di Padova, Via Marzolo 8, 35131, Padova, Italy
} 
\address{$\dag$ INFN, Sezione di Padova, Via Marzolo 8, 35131, Padova,
Italy}


\begin{abstract}
  Relative fluctuations of observables in discrete stochastic systems are bounded at all times by the mean dynamical activity in the system, quantified by the mean number of jumps.
  This constitutes a kinetic uncertainty relation that is fundamentally different from the thermodynamic uncertainty relation recently discussed in the literature. The thermodynamic constraint is more relevant close to equilibrium while the kinetic constraint is the limiting factor of the precision of a observables in regimes far from equilibrium. This is visualized for paradigmatic simple systems and with an example of molecular motor dynamics.
  Our approach is based on the recent fluctuation response inequality by Dechant and Sasa [arXiv:1804.08250] and can be applied to generic Markov jump systems, which describe a wide class of phenomena and observables, including the irreversible predator-prey dynamics that we use as an illustration.
\end{abstract}



\maketitle

\paragraph*{Introduction --}
The characterization of dynamical fluctuations is one of the main themes of nonequilibrium physics~\cite{\refFT,\refFDT,\refNEI,hat01,bod04,mae08b,che13,ber15,pol15,maes_book}.
Works on this subject led, for instance, to fluctuation relations~\cite{\refFT}, fluctuation-response relations~\cite{\refFDT}, and recently nonequilibrium inequalities~\cite{\refNEI}.
The latter branch was sparked by the discovery of a thermodynamic uncertainty relation (TUR)~\cite{bar15}.
The TUR states that the variance of a current in a stochastic system must be above a lower bound related to the entropy produced in the whole system. It was first proven, with the theory of large-deviations, in steady states~\cite{gin16} also for finite times~\cite{hor17}. A consequence of the TUR is that dissipation is a price to pay for accessing the possibility of a large signal-to-noise ratio. This does not guarantee that such high performance can be reached, as there could be other limiting constraints. 

Several results suggest that entropy production is not sufficient for characterizing the physics of systems far from equilibrium~\cite{maes_book}, as it is the case for life processes~\cite{bai18}. Dissipation determines the response of equilibrium systems but far from equilibrium we need to take into account also non-dissipative aspects~\cite{bai09,fal16b,yol17}. The volume of transitions in a system, regardless of the entropy associated with their occurrence, is an example of non-dissipative quantity. This dynamical activity~\cite{lec05,mer05,lec07,gar09,maes_book} may be the number of jumps between states in discrete systems, which are those we will considered below. We are thus talking of {\em kinetic} aspects, sometimes named {\em frenetic} aspects~\cite{mae17,maes_book}.

Recently, Dechant and Sasa (DS) put forward a flexible formalism~\cite{DS18} (recalled briefly below) that yields a general fluctuation-response inequality. Its generality stems also from the key role played there by a Kullback-Leibler divergence, a measure that is clearly not limited to the description of nonequilibrium systems. With DS approaches~\cite{dec18,DS18} it is possible to prove the TUR, both in the steady state and in transient regimes.
Also the derivation of our main result starts from the DS fluctuation-response inequality~\cite{DS18}.

In this paper we introduce a {\em kinetic uncertainty relation} (KUR) for Markov jump systems in continuous time~\cite{gardiner}, which describe a wide range of systems (molecules hopping between states, chemical reactions, demographic dynamics, etc.).
This inequality limits observable fluctuations from an angle totally distinct from the constraint of the TUR and is expressed as a function of time and of generic observables (with finite average). It embodies previous inequalities for nonequilibrium steady states~\cite{pie16a,gar17} (see also~\cite{chi18}) and is easily applicable to any dynamics without thermodynamic interpretation.

\paragraph*{Formula --}
The DS inequality~\cite{DS18} involves the response of a quantity $\O$ to a generic variation in the system conditions, which we parametrize by $\alpha$. This brings the probability measure $P(\omega)$ to a new value $P^\alpha(\omega)$, where $\omega$ may denote an instantaneous state, a full trajectory, etc., and is defined in a space $\Omega$ where both $0<P(\omega)<\infty$ and $0<P^\alpha(\omega)<\infty$. These measures are normalized ($\int_\Omega d\omega P(\omega) = 1$ and $\int_\Omega d\omega P^\alpha(\omega) = 1$) and give rise to expectations $\mean{\O}=\int_\Omega d\omega P(\omega) \O(\omega)$ and $\mean{\O}^\alpha=\int_\Omega d\omega P^\alpha(\omega) \O(\omega)$. A first DS nonequilibrium inequality reads
\begin{equation}
  [\mean{\O}^\alpha - \mean{\O}]^2 \le 2 \mean{\Delta \O^2}^\alpha \KL^\alpha
  \label{DS1}
\end{equation}
where $\mean{\Delta \O^2}^\alpha = \mean{\O^2}^\alpha - (\mean{\O}^\alpha)^2$ is the variance of $\O$ with respect to the measure $P^\alpha$ and
\begin{equation}
\KL^\alpha = \int_\Omega d\omega P^\alpha(\omega) \ln \frac {P^\alpha(\omega)}{P(\omega)}
\end{equation}
is the relative entropy, or Kullback-Leibler divergence between the two probabilities.
For a small perturbation, $P^\alpha = P + \alpha \rho + O(\alpha^2)$ and this relative entropy may be written as $\KL = \alpha^2 C +  O(\alpha^3)$ with ``cost function'' $C$, where the prefactor $\alpha^2$ takes into account the leading term in the expansion of $\KL$.
Similarly, one finds $\mean{\Delta \O^2}^\alpha \simeq \mean{\Delta \O^2} + O(\alpha)$. By defining the susceptibility
\begin{equation}
\chi_{\O} = \left. \frac{\partial \mean{\O}^\alpha}{\partial \alpha}\right|_{\alpha=0}
\end{equation}
we may restate DS' result by picking up the leading factors $\sim \alpha^2$ from both sides of the inequality (\ref{DS1}),
\begin{equation}
  \label{DS2}
\chi_{\O}^2 \le 2 \mean{\Delta \O^2} C\,.
\end{equation}
Let us now turn to an interesting descendant of this inequality.

Our application of this approach is for Markov jump processes, namely systems with discrete states $i,j,$ etc., and evolving in continuous time, with transition rates $k_{i j}$ denoting the probability per unit time to jump to $j$ if the system is in $i$. Let us also define the escape rate of a state $i$ as $\lambda_i = \sum_j k_{i j}$. We are interested in studying a quantity $\O(t)$ evolving from time $t_0=0$ to time $t$ according to this dynamics. Thus, averages as $\mean{\O}_t$ denote expectations where each $\omega=\{i(s)|0\le s\le t\}$ is a trajectory and $P(\omega)$ is its path probability taking into account also the initial state density. The process does not need to be in a steady state, it may also be experiencing a transient phase.

Perhaps surprisingly, a simple global rescaling of rates $k_{i j}\to k^\alpha_{i j} = (1+\alpha) k_{i j}$ (hence $\lambda^\alpha_i = (1+\alpha)\lambda_i$) yields interesting results. A $\alpha$-modification may not correspond to a true physical perturbation~\cite{DS18}, in general it may be a virtual change.
Indeed, the rescaling of rates is equivalent to a global change in pace of the system and leads naturally to perturbed quantities that are just unperturbed ones evaluated at longer times if $\alpha\gtrsim 0$.
In particular, $\mean{\O}_t^\alpha = \mean{\O}_{t+\alpha t}$ and thus
\begin{equation}
\chi_{\O}(t) = t \frac {d\mean{\O}_t}{d t} \equiv t \mean{\dot \O}_t
\end{equation}
A key feature of this $\alpha$-perturbation is its cost function,
\begin{align}
  \label{C}
  C(t) & = \frac 1 2 \mean{\caK}_t\,,  \\
  \label{fre}
  \mean{\caK}_t  & \equiv \sum_{i\ne j}\mean{n_{i j}}_t
\end{align}
where the sum is over is the mean number of jumps $\mean{n_{i j}}_t$ between any ordered pair $(i,j)$ of states. 
The expectation $\mean{\caK}_t$ (whose derivation is detailed before the conclusions) is a measure of how on average the system has been active during the period $[0,t]$, i.e.~it is the mean dynamical activity of the system integrated over the period $[0,t]$.
Thus, the inequality (\ref{DS2}) generates a KUR,
\begin{equation}
  \label{KUR}
\frac{\left(t \mean{\dot \O}_t\right)^2}{\mean{\Delta \O^2}_t} \le \mean{\caK}_t
\end{equation}
In a steady state this equation matches previous formulas for currents~\cite{pie16a} and counting observables~\cite{gar17}. 
By defining the {\em precision} of $\O$ at time $t$ as 
\begin{equation}
g_\O(t) = \frac{\mean{\dot \O}_t^2}{\mean{\Delta \O^2}_t/t}\,,
\end{equation}
and a mean frequency of jumps $\kappab(t) = \mean{\caK}_t/t$, we may rewrite the inequality as an upper kinetic bound on the precision,
\begin{equation}
  \label{KUR2}
g_\O(t) \le  \kappab(t)\,.
\end{equation}
For better graphical comparisons, in some examples we will plot $g_\O(t) / \kappab(t) \le 1$, which quantifies how close is the inequality (\ref{KUR2}) to being saturated.
Similarly, denoting by $\mean{\Sigma}_t$ the total entropy (in units with the Boltzmann constant $k_B=1$) produced in the time interval $[0,t]$ and by $\sigmab(t) = \mean{\Sigma}_t/t$ the mean rate of entropy production, we quote the (time-dependent) TUR as
\begin{align}
  \label{TUR}
  \frac{\left(t \mean{\dot \O}_t\right)^2}{\mean{\Delta \O^2}_t} \le \frac 1 2 \mean{\Sigma}_t
  \,,\\
  \label{TUR2}
  g_\O(t) \le  \frac 1 2\sigmab(t) \,,
\end{align}
which was expressed in this form by DS~\cite{dec18}.
In the following,  we drop the time dependence from symbols in case we deal with steady state averages.

\paragraph*{Examples --}

\begin{figure}[!t]
  \begin{center}
    \includegraphics[width=.95\columnwidth]{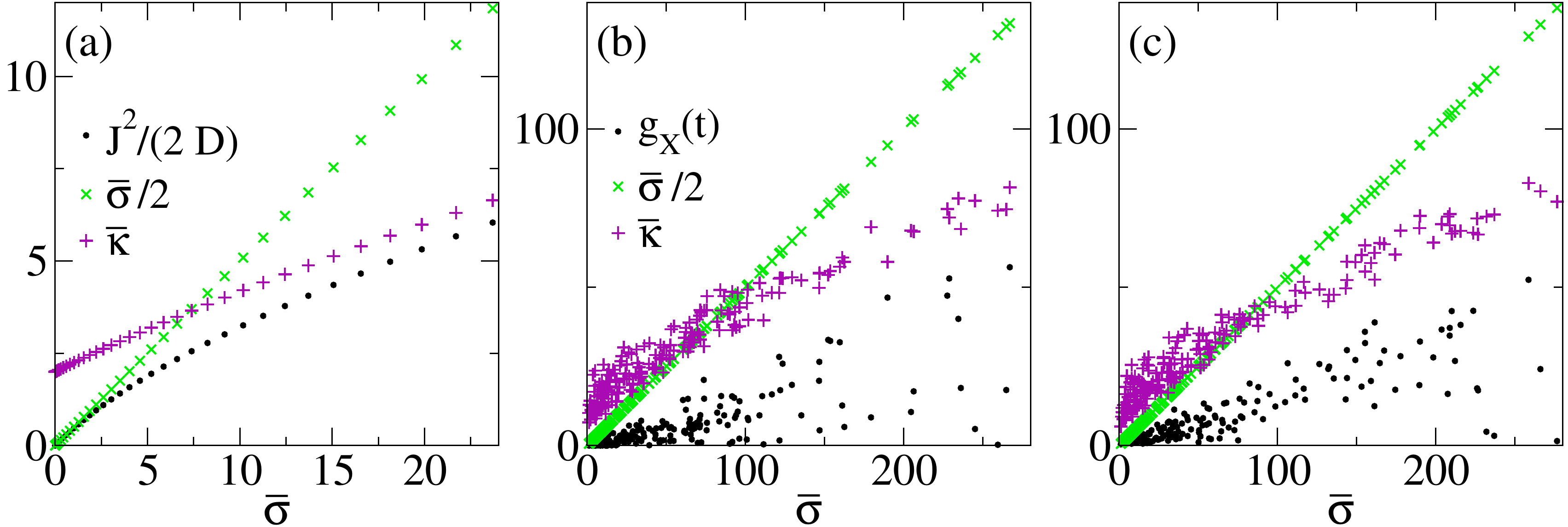}
    \end{center}
  \caption{Plots visualizing the TUR and KUR for different systems, as a function of the entropy production rate $\sigmab$, which quantifies the distance from equilibrium.
    (a) The precision $g_\O$ of the displacement for a one-dimensional biased random walk is bounded from above by  $\sigmab/2$ (TUR) and  $\kappab$ (KUR). (b) The same plot for a current integrated over a finite time $t$ in a random network with $N=4$ nodes and (c) for an observable depending nonlinearly on countings (see the text for more details).  }
\label{fig:random}
\end{figure}

One may wonder whether kinetic constraints are useful in thermodynamic systems~\cite{pie16a}. In fact, close to equilibrium there is a finite activity (the system jumps also in equilibrium) while entropy production tends to zero. Hence, around equilibrium the TUR brings certainly a tighter constraint on a current precision than the KUR~\cite{pie16a}. However, the standard example of a one-dimensional biased random walker shows immediately that far from equilibrium the ranking can be reversed. A force $F$ (normalized by unit displacement over $k_B T$) is applied toward positive values, and a walker jumps from $i$ to $i+1$ with rate $p=e^{F/2}$ or to $i-1$ with rate $q=e^{-F/2}$ so that local detailed balance is fulfilled: $\log p/q = F$ corresponds to the entropy added to the environment for $i\to i+1$, and the average rate of entropy production is
$\sigmab = \mean{\Sigma}_t/t = (p-q) F$. Clearly this value exceeds the mean jumping rate $\kappab = (p+q)$ for sufficiently large forces, where the KUR becomes the limiting factor for the precision. The latter, for $\O$ equal to the displacement at time $t$ from the initial position, is given by  $g_\O=J^2/(2 D)$ with average current $J=\mean{\dot \O} = (p-q)$ and diffusion constant $D=p+q=\kappab$.
In Fig.~\ref{fig:random}(a) we show the precision together with $\sigmab/2$ and $\kappab$ as a function of the mean dissipation rate $\sigmab$ (points parametrized by increasing $F$). The figure visualizes how $g_\O$ is first limited from above by the entropic constraint and then, getting farther from equilibrium by increasing $\sigmab$, it becomes bounded by the frenetic limit $g\le \kappab$.

The generality of the nonequilibrium frenetic bounds on the precision of a stochastic current is illustrated with a second example, in which we focus on finite times. Let us consider networks of $N$ states fully connected by transition rates
$k_{i j} =\exp(s_{i j}/2)$ and reverse $k_{j i}=\exp(-s_{i j}/2)$, with $s_{i j}$ drawn randomly from the interval $[-\Delta S_{\max},\Delta S_{\max}]$. The entropy production in a trajectory visiting states $\{i_0,i_1,\ldots,i_M\}$ is thus the sum $\Sigma(t) = \sum_{m=1}^M s_{i_{m-1} i_m}$ plus boundary terms that do not matter after averaging in the steady state.  Due to the randomization, some networks dissipate on average more than others. Using for instance $N=4$ and $\Delta S_{\max}=5$, we generate a wide spectrum of $\sigmab$ values.

We first choose to observe the current $\O(t) = n_{x y}(t) - n_{y x}(t)$ over the single bond $(x,y)$ with largest $s_{xy} =\max_{i,j}{s_{ij}}$ in each network. We thus count the jumps from $x$ to $y$ minus those from $y$ to $x$ up to time $t$, with $t=1 / \min_{i\ne j}{k_{i j}}$ small enough to not sample the steady state value for $g_X(t)$.
In Fig.~\ref{fig:random}(b) each point represents an average for a given random network. We see again that the time-dependent TUR~\cite{hor17,DS18} ($g_X(t)<\sigmab/2$) limits the current precision near equilibrium while the KUR ($g_X(t)<\kappab$) is the limiting factor far from equilibrium.

The observable $X$, however, needs not to be a current or an absolute counting. To show this, let us consider a weird nonlinear combination of local activity and global entropy production, $\O(t) = [(n_{x y}(t) + n_{y x}(t)) \Sigma(t)]^{1/5}$.
In Fig.~\ref{fig:random}(c) we see that also this observable obeys the KUR and the TUR~\cite{DS18}.
Here, for simplicity, we have always started from steady state conditions. The KUR would also hold by starting trajectories from generic conditions. The same is true for the TUR, if the Shannon entropy change is also included in $\Sigma(t)$~\cite{dec18}.

A molecular motor model confirms that the KUR is relevent in real nonequilibrium regimes. The model~\cite{lau07,lac08} is based on parameters extracted from experimental data and describes the motion of a kinesin molecule on a microtubule while pulling against an external force $F<0$.
Kinesin can be either in state $A$ (both arms on the tubule) or state $B$ (one arm raised). The transition from $A$ to $B$ can be activated either by thermal fluctuations or by ATP consumption.
We compute analytically $\sigmab$,  $\kappab$, and  the scaled cumulants of the motor displacement (mean velocity $J$ and second scaled cumulant $C_2$) by means of large-deviation theory~\cite{lac08,gar09}, see the Appendix for more details on this method and on the kinesin model. As before, the precision is $g_X=J^2/C_2$. 

\begin{figure}[!t]
  \begin{center}
    \includegraphics[width=.68\columnwidth]{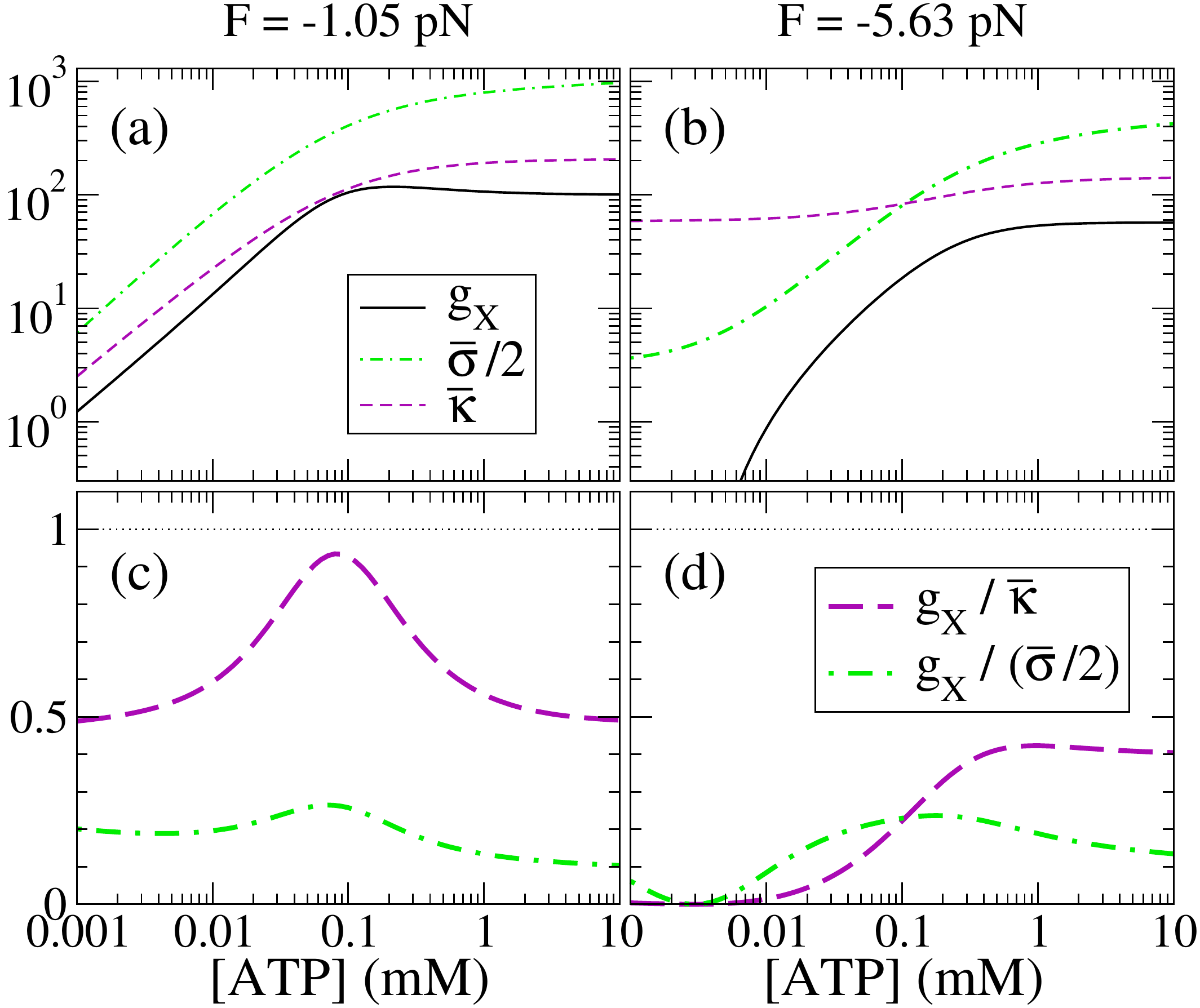} 
    \end{center}
  \caption{Model of kinesin pulling against a force $F$: left panels for $-F=1.05$pN, right panels for $-F=5.63$pN.
    (a), (b) Precision $g_X$ of the molecular motor displacement in the steady state as a function of ATP concentration, together with half mean dissipation $\sigmab/2$ and mean jumping rate $\kappab$. (c), (d) Ratio of the precision with these quantities.
    All plots highlight that in physiological conditions $1 \textrm{mM}\lesssim \textrm{[ATP]}\lesssim 10 \textrm{mM}$ it is the KUR that determines the upper limit of precision.
}
\label{fig:kin}
\end{figure}

In typical in vivo conditions ($-F\approx$ pN, ATP concentration  $1 \textrm{mM}\lesssim \textrm{[ATP]}\lesssim 10 \textrm{mM}$), in Fig.~\ref{fig:kin} we see that $\kappab<\sigmab/2$, i.e., it is the kinetic constraint that puts a ceiling on the precision of kinesin motion. At the smaller force [Fig.~\ref{fig:kin}(a),(c)], there is a regime of almost optimized precision at around [ATP]$=0.1$mM. While a lower precision seems normal at small fuel concentrations, it is perhaps more surprising when the environment furnishes more resources. However, this should not be considered unusual in life processes~\cite{bai18}. A nontrivial combination of dissipation with frenetic aspects may underlie this behavior for systems far from equilibrium.

The KUR is valid also for non-thermodynamic systems. We may for example consider processes without microscopic reversibility (some $k_{i j}\ne 0$ while $k_{j i}= 0$), where the TUR cannot be applied. The following example of population dynamics falls in this category.

\begin{figure}[!t]
  \begin{center}
  \includegraphics[width=.68\columnwidth]{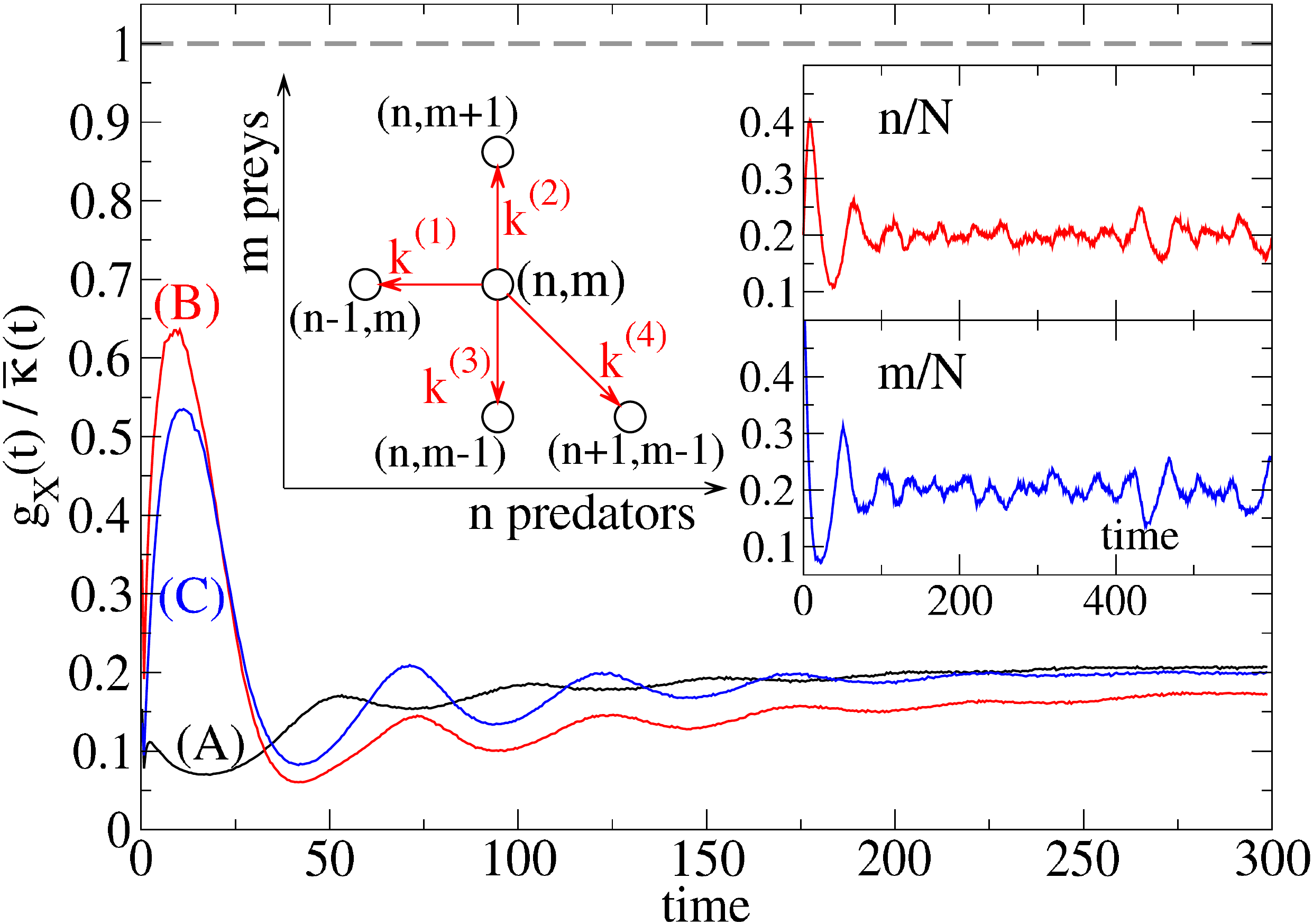} 
    \end{center}
  \caption{For the prey-predator model, we plot the KUR precision ratio $g_\O(t) / \kappab(t)$ for the counting $\O$ of predator deaths up to time $t$. Curves are for the three initial conditions discussed in the text ($N=3200$, $b=0.1$, $d_1=0.1$, $d_2=0$, $p_1=0.25$, $p_2=0.05$, curves from averages over $10^4$ trajectories). The horizontal dashed line highlights that $g_\O(t) / \kappab(t) \le 1$.
    The insets show a scheme of the transition rates of the model, and a relaxation trajectory from initial state of condition (C) toward a stochastic oscillatory regime.
}
\label{fig:sa}
\end{figure}

Stochastic equations are routinely applied in studies of population dynamics, of which a simple model is the predator-prey dynamics that gives rise quasi-periodic oscillations (due to {\em stochastic amplification}~\cite{mck05}). The system is described by the number of predators ($n$ individuals of kind $A$) and preys ($m$ individuals of kind $B$) in a niche allowing at most $N$ individuals, i.e., $n+m\le N$. Working in the context of urn models~\cite{mck05}, where also empty slots ($E$) are considered, the rates that describe the escape from a state $i=(n,m)$ are built upon microscopic processes
(birth $BE \xrightarrow[]{b} BB$,
death $A \xrightarrow[]{d_1} E$, $B \xrightarrow[]{d_2} E$ and
predation  $AB \xrightarrow[]{p_1} AA$, $AB \xrightarrow[]{p_2} AE$).
They become
$k^{(1)}=d_1 n$,
$k^{(2)}=2 b \frac m N (N-n-m)$,
$k^{(3)}=2 p_2 \frac {n m} N  + d_2 m$, and
$k^{(4)}=2 p_1 \frac {n m} N$,
see the scheme in the inset of Fig.~\ref{fig:sa}.

As an observable, in this demographic model we consider the number $\O(t)$ of predators' deaths (clearly originated by an irreversible process) up to time $t$, related to the transition $1$ with rates $k^{(1)}(n,m)$. The system is simulated with a Gillespie algorithm~\cite{gil77} and is released at time zero from a given initial condition: (A) in steady state, (B) with an abundance of predators ($n/N = 3 m/N = 0.6$), or (C) with an abundance of preys ($m/N = 3 n/N = 0.6$). In the large inset of Fig.~\ref{fig:sa} we see an example of relaxation toward the regime with stochastic oscillations, for case (C). We collect the statistics of $\O$ in this transient phase. Fig.~\ref{fig:sa} shows that in all cases the ratio $g_\O(t)/\kappab(t)\le 1$, as expected.

To compute the mean jumping rate $\kappab(t)$ in the simulation it is sufficient to record the total number of transitions in $[0,t]$ and average this value over many realizations of the process. We stress that in this system it is impossible to talk about irreversibility in the sense of ratios of reciprocal rates, $k_{i j}/k_{j i}$. Indeed such ratio is not defined for transitions $1$
and $4$, which do not have allowed reversals. Conversely, it seems a natural and easy procedure to count events to determine the mean dynamical activity.

\paragraph*{Derivation --}
To prove (\ref{C})-(\ref{fre}) we need to find the Kullback-Leibler divergence generated by the rescaling, to order $\alpha^2$, namely the expectation $\mean{\ln P^\alpha(\omega) / P(\omega)}_t^\alpha$.
For a given trajectory $\omega = \{i(s)| 0\le s \le t\}$ of a jump process, $i(s)$ is a piece-wise constant function of time which performs $M$ jumps between states $\{i_0,\ldots,i_M\}$. The ratio of path probabilities reads
\begin{align}
  \frac{ P^\alpha(\omega)}{P(\omega)}
  & = 
  \exp\left[ \int_0^t (\lambda_{i(s)} - \lambda^\alpha_{i(s)}) ds \right]
  \prod_{m=1}^M\frac {k^\alpha_{i_m,i_{m+1}}} {k_{i_m,i_{m+1}}} \nonumber\\
  & =  \exp\left[ -\alpha \int_0^t \lambda_{i(s)} ds \right] 
  \prod_{i\ne j}(1+\alpha)^{n_{i j}} \nonumber
\end{align}
The average of its log,
\begin{align}
  \label{logPP}
  \Mean{\ln \frac{P^\alpha(\omega)}{P(\omega)}}_t^\alpha
  =  -\alpha \Mean{\int_0^t  \lambda_{i(s)} ds}_t^\alpha
  + \ln(1+\alpha)\sum_{i\ne j}\mean{n_{i j}}^{\alpha}_t
\end{align}
contains a first term that, with the (time-dependent) perturbed state density $\rho^\alpha_i(s)$, can be rewritten as
\begin{align}
  &\Mean{\int_0^t  \lambda_{i(s)} ds}_t^\alpha
  =
  \frac{1}{1+\alpha}\Mean{\int_0^t  \lambda^\alpha_{i(s)} ds}^\alpha_t \nonumber\\
  &=
  \frac{1}{1+\alpha} \int_0^t \sum_{i\ne j} \rho^\alpha_i(s) k^\alpha_{i j} ds \nonumber\\
  &=
  \frac{1}{1+\alpha}\int_0^t \sum_{i\ne j} \frac{d\mean{n_{i j}}^\alpha_s}{ds} ds
  =
  \frac{1}{1+\alpha} \sum_{i\ne j} \mean{n_{i j}}^\alpha_t\nonumber
\end{align}
This in (\ref{logPP}) yields a prefactor $\ln(1+\alpha)-\alpha/(1+\alpha) \sim\alpha^2/2$ for $\mean{n_{i j}}_t^\alpha$, allowing us to replace it by $\mean{n_{i j}}_t$ to leading order. Thus, we finally have the Kullback-Leibler divergence leading to (\ref{C})-(\ref{fre}),
\begin{align}
  \label{almost}
  \Mean{\ln \frac{P^\alpha(\omega)}{P(\omega)}}_t^\alpha
  &= \frac{\alpha^2} 2 \sum_{i\ne j} \mean{n_{i j}}_t + O(\alpha^3)
\end{align}

\paragraph*{Conclusions --}
The kinetic constraint described in this work is a further characterization of fluctuations in stochastic systems. Such time-dependent uncertainty relation for a fluctuating quantity takes into account the mean dynamical activity of the whole system and includes steady state versions~\cite{pie16a,gar17} as special cases. The universality of the DS inequality~\cite{DS18} for us yields a formula for generic regimes (transient, oscillatory, steady, etc.) and generic observables. The DS inequalities~\cite{dec18,DS18} include also the TUR as another special case. The same approach leads to other inequalities~\cite{DS18,D18} not characterized by entropy production or dynamical activity as in TUR and KUR, respectively, but by other cost functions.

For thermodynamic systems, the KUR is complementary to the uncertainty relation based on thermodynamic considerations and focusing on dissipation. We have shown examples in which, by getting farther and farther from equilibrium dynamics, the kinetic constraints  become more relevant than thermodynamic ones in limiting the precision of a given process. A model of kinesin suggests that the dispersion in the motion of this molecular motor, in physiological conditions, cannot be arbitrarily small due to the KUR.

Moreover, the KUR is readily computed also in irreversible systems as those often modeled by discrete stochastic processes. With an example of predator-prey dynamics, we have illustrated how the KUR puts a limit on the maximum precision of observables.

\ack  We thank Gianmaria Falasco, Juan Garrahan, Christian Maes, and Patrick Pietzonka for useful discussions and constructive comments on the manuscript.
  MB acknowledges support from Progetto di Ricerca Dipartimentale BIRD173122/17.

\section*{Appendix}

We recall the details of the molecular motor model~\cite{lau07,lac08} used in the main text and of the analytical technique for computing average quantities from a scaled cumulant generating function in the context of large deviation theory~\cite{lac08,gar09}.

\begin{figure}[!tb]
  \begin{center}
    \includegraphics[width=8cm]{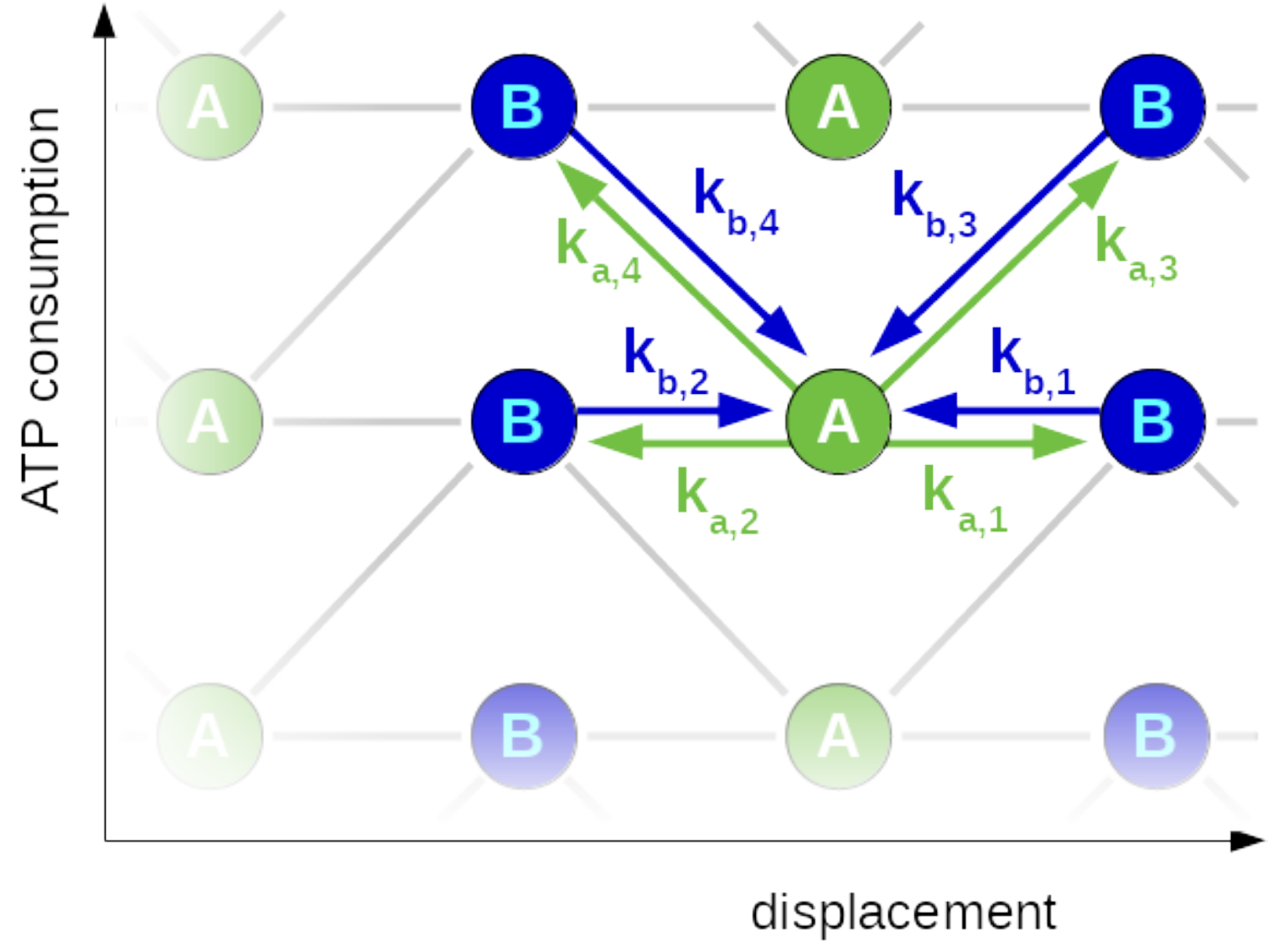} 
    \end{center}
  \caption{Sketch of possible transitions and relative transition rates in the kinesin model.    }
\label{fig:sch}
\end{figure}

Kinesin walks on a microtubule, essentially in one dimension with half steps of size $d=4$nm, and can be in one of two kinds of state, $A$ (its legs are on the tubule) and $B$ (one leg is raised).
The full state space is infinite and characterized by the number $n$ of half steps of the motor and by the number $m$ of ATP consumed, after assuming for example $i=(0,0)$ to be of kind $A$.
A jump from a state $i=(n,m)$ is to a neighbor of the opposite kind via one of four different transitions. For kind $A$ these are: (1) right step with no ATP consumed, (2) left step with no ATP consumed, (3) right step with ATP consumed (which is the typical motor direction), (4) left step with ATP consumed.
The opposite transitions from state $B$  are also possible, see the scheme in Fig.~\ref{fig:sch}.
In a notation consistent with the sketched transitions, we write their rates as
$\vec k_a = (k_{a,1},k_{a,2},k_{a,3},k_{a,4})$, $\vec k_b = (k_{b,1},k_{b,2},k_{b,3},k_{b,4})$.
Thus, in total, there is a rate $\lambda_a = \vec k_a \cdot \vec 1$ to escape from $A$ and $\lambda_b = \vec k_b \cdot \vec 1$ from $B$, where $\vec 1=(1,1,1,1)$.

Kinesin is usually pulling cargoes with a force $F\approx $ pN. 
For a scaled force $f = F d / (k_B T)$ with sign $f<0$ if $F$ is opposite to the motion (it pulls on the left), rates read
\begin{align}
  k_{a,1} & = \omega e^{-\epsilon + \theta_a^+ f}  \,, \qquad&
  k_{b,1} & = \omega e^{-\theta_b^- f}   \,, \nonumber\\
  k_{a,2} & = \omega' e^{-\epsilon - \theta_a^- f} \,, \qquad&
  k_{b,2} & = \omega' e^{\theta_b^+ f} \,, \nonumber\\
  k_{a,3} & = \alpha e^{-\epsilon + \theta_a^+ f} k_0[\textrm{ATP}] \,, \qquad&
  k_{b,3} & = \alpha e^{-\theta_b^- f} \,, \nonumber\\
  k_{a,4} & = \alpha' e^{-\epsilon - \theta_a^- f} k_0[\textrm{ATP}]\,, \qquad&
  k_{b,4} & = \alpha' e^{\theta_b^+ f} \,, \nonumber
\end{align}
with values
\begin{align}
\theta_a^+ = 0.25\,,\quad
\theta_a^- = 1.83\,,\quad
\theta_b^+ = 0.08\,,\quad
\theta_b^- = -0.16\,,\nonumber
\end{align}
and $k_0 = 1.4 \times 10^5\mu\rm{M}^{-1}$, $\epsilon=10.81$,
$\alpha = 0.57 \rm{s}^{-1}$,
$\alpha' = 1.3 \times 10^{-6}\rm{s}^{-1}$,
$\omega= 3.4 \rm{s}^{-1}$,
$\omega'= 108.15 \rm{s}^{-1}$.
See Ref.~\cite{lau07} for all the details on these parameters.

The generator $\mathbf{M}$ of the dynamics is the matrix entering in the master equation $\partial_t  p = \mathbf{M}  p$, where $p = (p_A,p_B)$ is the probability at time $t$ to find the motor in $A$ or $B$.
To obtain the cumulants of an observable $X$, one first ``tilts'' the generator with exponentials suitably coupled to the transitions. Let us define $\vec v_i = (e^{\gamma_{i,1} s},e^{\gamma_{i,2} s},e^{\gamma_{i,3} s},e^{\gamma_{i,4} s})$ where $\gamma$'s will define the observable $X$. With this definition, the tilted generator is written in a compact notation as
\begin{displaymath}
\mathbf{M}_X(s) =
\left( \begin{array}{cc}
-\lambda_a & \vec k_b \cdot \vec v_b \\
\vec k_a \cdot \vec v_a & -\lambda_b
\end{array} \right)
\end{displaymath}
Its eigenvalue $\Lambda(s)$ (of the two) for which $\Lambda(0)=0$ (i.e. the eigenvalue corresponding to the steady state of $\mathbf{M} = \mathbf{M}_X(0)$) is the scaled cumulant generating function of $X$, namely the function determining its scaled cumulants in the limit of long time.
For instance, the scaled average is
\begin{equation}
J=\lim_{t\to\infty}\mean{X}_t / t = \left.\partial_s \Lambda_X(s)\right|_{s=0} \nonumber
\end{equation}
and the scaled variance is
\begin{equation}
  C_2=\lim_{t\to\infty}[\mean{X^2}_t - \mean{X}_t^2] / t = \left.\partial^2_s \Lambda_X(s)\right|_{s=0}
   \nonumber
\end{equation}
To obtain the mean dynamical activity $\kappab$ one uses $\vec \gamma_a=\vec \gamma_b=\vec 1$. For the entropy production $\sigmab$ (with Boltzmann constant $k_B=1$) the suitable
$\vec \gamma_a=-\vec \gamma_b=(\ln k_{a,1}/k_{b,1},\ldots,\ln k_{a,4}/k_{b,4})$.
For an observable $X$ equal to the displacement, in this model we have $\vec \gamma_a=-\vec \gamma_b=(d,-d,d,-d)$.
The solutions plotted in the main text are obtained via Mathematica and here we do not report their long explicit formulae.

\section*{References}


\providecommand{\newblock}{}

\end{document}